\documentstyle[aps,twocolumn,prb,epsf]{revtex}

\begin{document}

\newcommand{\be}{\begin{equation}}
\newcommand{\ee}{\end{equation}}
\newcommand{\bq}{\begin{eqnarray}}
\newcommand{\eq}{\end{eqnarray}}

\draft

\twocolumn[\hsize\textwidth\columnwidth\hsize\csname @twocolumnfalse\endcsname

\draft

\title{Role of Orbital Degeneracy in Double Exchange Systems}

\author{M. S. Laad$^1$, L. Craco$^2$ and E. M\"uller-Hartmann$^1$}
\address{${}^1$Institut f\"ur Theoretische Physik, Universit\"at zu K\"oln, 
77 Z\"ulpicher Strasse, D-50937 K\"oln, Germany \\
${}^2$Instituto de F\'{\i}sica Gleb Wataghin - UNICAMP, C.P. 6165, 13083-970 
Campinas - SP, Brazil} 

\date{\today}

\maketitle

\widetext

\begin{abstract}
We investigate the role of orbital degeneracy in the double exchange (DE)
model. In the $J_{H}\rightarrow\infty$ limit, an effective generalized 
``Hubbard'' model incorporating orbital pseudospin degrees of freedom 
is derived. The model possesses an exact solution in one- and in infinite 
dimensions. In 1D, the metallic phase off ``half-filling'' is a Luttinger 
liquid with pseudospin-charge separation. Using the $d=\infty$ solution 
for our effective model, we show how many experimental observations for 
the well-doped ($x\simeq 0.3$) three-dimensional manganites 
$La_{1-x}Sr_{x}MnO_{3}$ can be qualitatively explained by invoking
the role of orbital degeneracy in the DE model.  
\end{abstract}
     
\pacs{PACS numbers: 71.28+d,71.30+h,72.10-d}

]

\narrowtext

Colossal magnetoresistance (CMR) materials are presently the focus of much
experimental and theoretical attention~\cite{[1]}. These materials are 
important technologically, owing to the huge decrease of resistivity in 
modest external magnetic fields ($\Delta R/R \simeq 100000$).

Early studies of the models for manganites~\cite{[2]} concentrated on the 
observed link between magnetic and transport properties and studied the 
``double exchange'' (DE) model. This simple model certainly has had a 
surprising degree of success in explaining the existence of the low-$T$ 
ferromagnetic metallic (FM) phase in doped $La_{1-x}Sr_{x}MnO_{3}$ with 
$x=0.3$. However, it has come to be subsequently realized that several 
features of even the high-$T$ phase at $x=0.3$, not to mention the 
existence of a very rich phase diagram~\cite{[3]}, are inexplicable 
within the framework of the pure DE model.  The insulating ``normal'' 
state above $T_{c}^{FM}$ at $x=0.33$ points to the existence of additional 
localizing mechanism/s; various candidates considered are (1) a strong 
coupling of carriers to phonon degrees of freedom~\cite{[4]}, (2) FM 
short-range order and/or Berry phase effects in the DE model, leading to 
fluctuations in the hopping and to carrier localization for sufficiently 
strong spin disorder~\cite{[5]}.  The phase diagram of 
$La_{1-x}Ca_{x}MnO_{3}$ is very rich with regions of ferromagnetic, 
antiferromagnetic and charge order as a function of $x$~\cite{[1]}.
Optical conductivity measurements and photoemission studies~\cite{[6]} have 
revealed the transfer of spectral weight over large energy scales, a 
characteristic of strongly correlated systems.  In addition, these studies 
indicate a small discontinuity in the momentum space occupation fn. 
$n({\bf k})$, showing up the strongly correlated nature of the FM metallic 
state. However, the low-$T$ electronic specific heat shows only a modest 
enhancement of about 2-3 times the bandstructure value~\cite{[7]}. The dc 
resistivity below $T_{c}^{FM}$ follows $\rho(T)=\rho_{0}+AT^{2}+BT^{9/2}$; 
the first and the last terms can be reconciled with the DE model~\cite{[7]}, 
but the $AT^{2}$ term with a large $A$ decreasing in an applied field, cannot.
Recently, Simpson {\it et al.}~\cite{[8]} have carried out a careful study 
of the optical response of $La_{0.7}A_{0.3}MnO_{3}$ ( A=Ca, Sr ) crystals, 
which are ferromagnetic metals at low-$T$.  
In contrast with earlier measurements
yielding optical masses very different from those extracted from specific 
heat data, these authors have found that the effective mass is similarly 
enhanced in both cases (it is shown~\cite{[8]} that the differences reported 
earlier were artifacts of surface effects).  Moreover, they extract the 
scattering rate from their data: the result is 
$\tau^{-1}(\omega) \simeq -C-D\omega^{2}$, completely consistent with the 
quadratic $T$-dependent component in the resistivity.  Raman scattering 
experiments reveal more confirmation of the importance of electronic
correlations below $T_{c}^{FM}$~\cite{[8]}; above $T_{c}^{FM}$, the 
lineshapes can be fit nicely with a constant, large damping, but below 
$T_{c}^{FM}$, good fits can only be obtained by assuming a frequency dependent 
$\Gamma(\omega)=\Gamma_{0}+\alpha\omega^{2}$, with $\alpha$ increasing 
with decreasing $T$, in full 
accord with the decrease of the dc resistivity.  Raman results probing the 
field dependence of the scattering rate from different channels also shows 
that the magnetoresistance correlates well with the change in the scattering
rate in the $B_{1g}$ channel in an applied field, but that the $A_{1g}$ 
scattering rate is almost independent of $B_{ext}$, suggesting an anisotropic
scattering rate. Finally, the dc Hall data is very anomalous and defies 
an explanation,~\cite{[9]} the normal part implying a large Fermi surface 
right upto the insulator-metal phase boundary, and the anomalous Hall constant
scaling like $[M(0)-M(T)]^{3/2}$ instead of  $[M(0)-M(T)]^{3}$ as expected 
from a DE model.  The above-mentioned facts force one to conclude the existence
of an additional scattering mechanism inducing a new low energy coherence 
scale in the FM metallic state.  

The important role of carrier lattice interactions is demonstrated clearly
by the isotope effect in $T_{c}^{FM}$~\cite{[10]}, which changes by $20 K$ on 
substitution of $O^{18}$ by $O^{16}$; such behavior requires a 
consideration of {\it dynamical} hole-lattice interactions 
(it is absent in the static limit).  Also important in this context 
are experiments which reveal small polaron transport in the
high-$T$ paramagnetic insulating regime (at $x=0.3$)~\cite{[11]}. Since 
$T_{c}^{FM}$ in the strong coupling limit of the DE model scales with the 
fermion bandwidth, an electronic bandwidth renormalized by carrier-phonon 
interaction might be expected to show up the isotope effect in $T_{c}^{FM}$.  

The regime of charge ordering is more or less unexplored, but seems to be
quite anomalous.  Recent experimental work has underscored the strong 
correlation between the magnetic, orbital and lattice degrees of freedom 
in the manganites.  In $La_{1-x}Sr_{x}MnO_{3}$, with $x=1/8$, 
charge-orbital ordering (COO) was observed by x-ray and neutron 
diffraction~\cite{[12]}, close to commensurate filling fraction of the
carriers.  This is accompanied by a reentrant transition to an insulating 
state.  Intriguingly, this COO state is {\it stabilized} by an
applied magnetic field $B_{ext}$, in contradiction to what the pure DE model
would suggest.  This is because, in the pure DE picture, an external magnetic
field would align the core spins, increase the carrier kinetic energy, and 
{\it weaken} the COO state.  As mentioned in~\cite{[12]}, this means that the
COO insulating state at low-$T$ is more fully magnetized than the
FM-metallic state.  In contrast to this, the charge order arising in the 
$Nd_{0.5}Sr_{0.5}MnO_{3}$ samples (which are antiferromagnetically ordered) is
weakened by applied fields, suggesting a deep connection between magnetic and
orbital degrees of freedom that cannot be accessed within the pure DE model.
  
  Clearly, in view of the above anomalous features, effects of orbital 
degeneracy and carrier-lattice interactions should be considered in
a model hamiltonian within a framework which treats their interplay with 
the magnetic degrees of freedom on an equal footing.

In this paper, we address the issue of the interplay between the magnetic 
and orbital degrees of freedom.  The problem has been a subject of 
investigation for many years, beginning with the Kugel-Khomskii~\cite{[13]} 
paper. In recent years, it has become clear that these effects are crucial 
for a proper understanding of doped transition metal (TM) oxides~\cite{[14]}.  
However, the full range of new interesting effects has by no means been 
accessed, and investigation on the role of orbital degeneracy in correlated 
systems is an active, open problem.   

  We start with a model that explicitly includes the effects of orbital 
degeneracy in the $e_{g}$ orbitals in manganites~\cite{[15]},

\bq
\nonumber
H & = & -t_{1}\sum_{<ij>}(a_{i\sigma}^{\dag}a_{j\sigma}+
b_{i\sigma}^{\dag}b_{j\sigma}+h.c) -t_{2}\sum_{<ij>}
(a_{i\sigma}^{\dag}b_{j\sigma}+h.c) \\ \nonumber
& + & U\sum_{i}(n_{ia\uparrow}n_{ia\downarrow}+
n_{ib\uparrow}n_{ib\downarrow}) + 
U_{ab}\sum_{i\sigma\sigma'}n_{ia\sigma}n_{ib\sigma'} \\ 
& - & J_{H}\sum_{i\sigma\sigma'}{\bf S_{i}^{c}.{\bf \sigma_{i}}}
(a_{i\sigma}^{\dag}a_{i\sigma'}+b_{i\sigma}^{\dag}b_{i\sigma'}) 
+ H_{J-T} \;,
\eq
where the $a$ and $b$ describe fermion annihilation operators in the doubly 
degenerate $e_{g}$ orbitals, $U_{ab}$ is the on-site, interorbital repulsion,
$U$ is the on-site, intraatomic Hubbard interaction, and $J_{H}$ is the 
(strong) Hund's rule coupling giving rise to the FM state as in the usual DE
model.  We have also included the polaronic effects via $H_{J-T}$, as described
later in the following. 
It should be mentioned that the structure of the hopping matrix elements 
corresponding to the realistic 3d perovskite structure is more complicated
than what we have assumed in the model above.  This should be crucial when 
one attempts to focus on the various observed orbitally ordered (with 
ferro- or antiferromagnetic order) phases in the global phase diagram.
However,  in the orbitally disordered ferromagnetic
metal phase occuring around $x=0.3$ that we are interested in here, we 
believe that a simplified modelling of the hopping matrix suffices, and a
more realistic choice will not qualitatively affect our conclusions.  In 
this context, we would like to mention that a similar simplification has
been assumed in~\cite{[15]} in a slightly different context.  Our main aim 
here is to demonstrate that orbital correlations, along with double exchange,
play a key role in the unified understanding of the various anomalies
observed in the FM state around $x=0.3$. 

  In the usual DE limit, we are in the regime of $U,J_{H} >> t_{1,2}$.  This
means that the carrier states having their spin antiparallel to the core spin
are projected out (we treat the core $t_{2g}$ spin classically, following
Furukawa), and so one is effectively dealing with spinless fermions, but with
one important difference; the fermions still carry an orbital index.  The 
usual DE model does not include these orbital degrees of freedom, and so cannot
access the interplay between magnetism and orbital ordering shown up in 
manganites.  We will show below how inclusion of orbital effects leads to 
new physics in the manganites, and how many intriguing observations have a 
natural explanation within our picture.

  In the DE limit, the effective hamiltonian is the same as eqn.~(1) without 
the spin index,
\bq
\nonumber 
H & = & -t_{1}\sum_{<ij>}\gamma_{ij}({\bf S})(a_{i}^{\dag}a_{j}
+b_{i}^{\dag}b_{j}) \\
& - & t_{2}\sum_{<ij>}\gamma_{ij}({\bf S})(a_{i}^{\dag}b_{j}+h.c) 
+ U_{ab}\sum_{i}n_{ia}n_{ib} \;,
\eq
where $\gamma_{ij}({\bf S})=\frac{<S_{c}>_{ij}+1/2}{2S+1}$~\cite{[5]} 
is the usual DE
factor which describes the fact that the $e_{g}$ holes increase their kinetic
energy by forcing the core $t_{2g}$ spins to align ferromagnetically.
At temperatures above the ferromagnetic $T_{c}^{FM}$, the FM short range order
would imply an effective model with hopping disorder.  Varma and 
M\"uller-Hartmann {\it et al}~\cite{[5]} have proposed that a sufficiently 
strong hopping disorder could lead to localization of carriers above 
$T_{c}^{FM}$; such ideas have, however, been challenged by Millis 
{\it et al}~\cite{[4]}, who argue for localization caused by strong JT 
effects at $x=0.3$.  However, no orbital (JT) order is observed at $x=0.3$ 
in the FM metallic state (the metallic state is a para-orbital Fermi liquid), 
leaving one to have to deal with coupled magnetic and orbital degrees of 
freedom.

Transforming to new pseudo-spin variables, $c_{\uparrow}=(a+b)/\sqrt{2}$,
$c_{\downarrow}=(a-b)/\sqrt{2}$ yields a generalized Hubbard model with 
pseudospin-dependent hoppings,
\bq
\nonumber
H_{eff} & = & -(t_{1}+t_{2})\sum_{<ij>}\gamma_{ij}({\bf S})
(c_{i\uparrow}^{\dag}c_{j\downarrow}+h.c) \\
& - & (t_{1}-t_{2})\sum_{<ij>}\gamma_{ij}({\bf S})
(c_{i\downarrow}^{\dag}c_{j\downarrow}+h.c) 
+ V\sum_{i}n_{i\uparrow}n_{i\downarrow} \;,
\eq
where we call $U_{ab}=V$.  In the translationally invariant FM metallic state,
$\gamma_{ij}({\bf S})=\gamma(M)$ where $M=<S_{i}^{z}>$ is the average core-spin
magnetization.  Redefining $(t_{1}+t_{2})\gamma(M)=t_{\uparrow}(M)$ and
$(t_{1}-t_{2})\gamma(M)=t_{\downarrow}(M)$, we have finally,
\bq
\nonumber
H_{eff} & = & -t_{\uparrow}(M)\sum_{<ij>}
(c_{i\uparrow}^{\dag}c_{j\uparrow}+h.c) \\
& - & t_{\downarrow}(M)\sum_{<ij>}(c_{i\downarrow}^{\dag}c_{j\downarrow}+h.c) 
+ V\sum_{i}n_{i\uparrow}n_{i\downarrow} \;.
\eq
  This describes an effective Hubbard-like model with magnetization dependent
hopping in {\it orbital} space.  This reflects the correlation between the
magnetic and orbital degrees of freedom that we are interested in here.  In
contrast to the usual one-band Hubbard model, the hamiltonian eqn.~(4) has a 
global symmetry $Z(2) \times U(1)$ in orbital space (it is mapped onto an 
effective $XXZ$ pseudospin model when $V>>t$, see below).  As we will show in 
what follows, the effective model, eqn.~(4) leads to a more interesting phase 
diagram than the one obtained from the usual one-band Hubbard model.

  In the strong coupling limit, $t\gamma(M) \ll V$, eqn.~(4) is mapped to an 
effective anisotropic pseudo-spin hamiltonian,
\bq
\nonumber
H_{eff} & = & 4\frac{(t_{1}+t_{2})^{2}\gamma^{2}(M)}{V}\sum_{<ij>}
\tau_{i}^{z}\tau_{j}^{z} \\
& + &  4\frac{(t_{1}^{2}-t_{2}^{2})\gamma^{2}(M)}{V}
\sum_{<ij>}(\tau_{i}^{x}\tau_{j}^{x}+\tau_{i}^{y}\tau_{j}^{y})
\eq  
  Our analysis below will be based both on eqs.~(4) and~(5).
To start with, we note the following features which have been exhaustively 
studied by many workers.~\cite{[16]}

(1) $t_{1}=t_{2}$ in eqn.~(1).  One observes from eqs.~(1)-(5) that 
the ``antisymmetric'' orbital contribution $(a-b)$ decouples from the problem,
but interacts with the ``symmetric'' fermion modes via a local coupling $V$.
In this limit, eqn.~(5) becomes a pure Ising model.  Eqn.~(4) reduces to the
well-studied Falicov-Kimball model~\cite{[16]}.  This has already been 
employed in connection with the anomalous metallic state in the FM 
manganites~\cite{[15]}. In this situation, the physics,
both at and off ``half-filling'', has been clear both in one- and in infinite
dimensions~\cite{[16]}. In 1D, the ground state is characterized by 
staggered orbital order at $T=0$, and at a finite $T=T_{c}>0$ in $D>1$.  
In $d=\infty$, the ground state is charge-ordered, while the paramagnetic 
metallic phase is a non-Fermi liquid, and is mapped onto the x-ray edge 
problem. In this case, the spectral fn. of the antisymmetric orbital mode, 
as well as the mixed orbital excitonic response fn, is singular at 
$T=0$~\cite{[16]}:
\be
Im G_{loc}^{\downarrow}(\omega) \simeq |\omega|^{-(1-\alpha)}
\ee
and
\be
\chi^{"}(\omega)=\int_{0}^{\infty}d\tau e^{i\omega\tau}\langle 
T_{\tau}c_{\downarrow}^{\dag}c_{\uparrow}(\tau);c_{\uparrow}^{\dag}
c_{\downarrow}(0)\rangle \simeq |\omega|^{-\beta} \;,
\ee
where $\alpha=(\delta/\pi)^{2}$, $\beta=(2\delta/\pi)-(\delta/\pi)^{2}$, with
$\delta=tan^{-1}(V/D)$ and $D$ the free bandwidth (with $V=0$).
Thus, this non-FL metallic phase is related to the occurence of a soft, 
local orbital excitonic mode in our approach.
  
The breakdown of FLT is caused by a ground state degeneracy
$[n_{i\downarrow},H_{eff}]=0$ for all $i$.  The exact $d=\infty$ solution 
shows also that the $n_{\downarrow}$ varies discontinuously as a function of
filling near ``half-filling''~\cite{[16]}, leading to orbital phase 
separation.  In our case, this corresponds to hole-rich regions 
(ferromagnetic by the underlying DE mechanism in $H_{eff}$, with $Mn^{4+}$, 
which is JT-inactive), coexisting with the hole-poor ($Mn^{3+}$, which is 
JT-active, with underlying AF coming from the remnant of the $x=0$ 
insulating state) regions. In this situation, the free energy of the model 
has a double minimum structure with $n_{\downarrow}=0,1$, leading to low 
energy multiparticle excitations and an incoherent ferromagnetic metal.  
Given a non-FL metal, the consequences of static disorder, intrinsic or 
otherwise, are drastic. 

(2) $t_{2}=0$ in eqn.~(1).  
The ``two-band'' model, eqn.~(1), reduces to the usual
Hubbard model (in orbital space).  This model has an exact solution in 1D,
as is known from the Lieb-Wu solution.  Away from "half-filling", it is a
non-FL metal with pseudospin-charge separation at low energy~\cite{[17]}.  
Due to the Mermin-Wagner theorem, finite temperature order is ruled out, 
but it is widely believed that the 2d version has an orbitally ordered 
ground state.  Much more is known in $d=\infty$~\cite{[16]}. The ground 
state, at and near "half-filling", is (in our model) charge ordered.  
The quantum para-orbital metallic phase obtained by suppressing the low-$T$ 
ordered state is a correlated Fermi liquid, with a dynamically generated 
low-energy coherence scale (related to orbital Kondo screening in our 
model~\cite{[16]}) which vanishes as the Mott insulating state (the
orbital Mott insulator in our model) is approached by $n\rightarrow 1$ or 
by $V>V_{c}$ at $n=1$.

  The most interesting situation, probably the one actually realized in 
practice, is the one where $t_{1} \ne t_{2}$.  Given the actual hopping 
parameters as in~\cite{[21]}, neither of the cases $t_{1}=t_{2}$ or 
$t_{2}=0$ is realistic. In this situation, let us consider 
the physical picture one would expect in $d=\infty$ (which is what we
use as an approximation for the 3d manganites).  In $d=\infty$, the lattice 
model is mapped on to a single-impurity Anderson model supplemented with a 
selfconsistency condition which describes the coupling of the arbitrarily
chosen (for a translationally invariant case) "impurity" site to the rest of 
the lattice.  At sufficiently high-$T$, the orbital moments are unquenched,
and the physics is effectively the same as that known from the high-$T$ limit
of the usual $d=\infty$ Hubbard model.  However, since $t_{\downarrow}(M)$ can
take on arbitrarily small values as $t_{1}-t_{2}$ is varied, the effective low
energy coherence scale can be driven quite low, and, in fact, in practice, can
be overshadowed by residual disorder effects in doped samples.  In the regime
where this is true, one is dealing with a two-component picture, with 
additional strong scattering provided by the orbital moments for 
$T>T_{K}^{orb}$.  This
additional incoherent scattering mechanism disappears at $T_{K}^{orb}$, below
which the orbital moments are quenched, leading to a coherent FL response.
The difference from the usual situation as encountered in the one-band Hubbard
model is that the core-spin thermal fluctuations and two-spinwave processes
also contribute.  One expects the transport properties to be additive with 
respect to scattering from orbital and spin fluctuations.  As we will discuss
below, aspects of the resistivity below $T_{c}^{FM}$ are naturally 
understandable in this picture. One would expect disorder or a modest 
magnetic field to cause dramatic changes in transport properties in this 
situation because the effective carrier bandwidth has already been narrowed 
down by orbital correlations, even in the para-orbital phase. At sufficiently 
low-$T$, there is an instability to a orbital-ordered state when the filling 
is close to commensurable values, in analogy with what is known about the 
Hubbard model in this limit.  

  We turn now to the effects of adding polaronic term via strong Jahn-Teller
coupling in the above formalism.  The modelling essentially duplicates that
employed by Millis {\it et al}~\cite{[4]}; 
here, the local lattice distortions which cause the Jahn-Teller splitting
transform as a two-fold degenerate representation of the cubic group which
couple to the electron as a traceless symmetric matrix.  The JT part is
\bq
\nonumber
H_{J-T} & = & \lambda\sum_{i}(Q_{2i}\tau_{i}^{z}+Q_{3i}\tau_{i}^{x}) 
+ K/2\sum_{i}(Q_{2i}^{2}+Q_{3i}^{2}) \\
& + & M_{0}/2\sum_{i}[(dQ_{2i}/dt)^{2}+(dQ_{3i}/dt)^{2}] \;,
\eq
where $Q_{2}=(2\delta Z-\delta X-\delta Y)/6^{1/2}$ and 
$Q_{3}=(\delta X-\delta Y)/2^{1/2}$, where $\lambda$ is the electron-phonon 
coupling, $K$ is the phonon stiffness, and $M_{0}$ the ionic mass.
We can define~\cite{[4]} the distortion vector ${\bf Q}$ by $Q_{x}=Q_{3}$ and 
$Q_{z}=Q_{2}$, so that $H_{JT}=\lambda\sum_{i}{\bf Q_{i}}.{\bf \tau_{i}}$
Further, we consider a fixed $|{\bf Q}|$, so that the distortions are modeled 
by a local classical spin.  In the adiabatic limit for the phonons, and for
strong coupling, the JT part acts like additional double exchange.  As $T$ 
increases, fluctuations in the $Q_{i}$ become important.  

Thus, an enormous simplification occurs:  the original two-band model is 
mapped onto a generalized orbital Hubbard model with magnetization dependent 
hoppings with an additional ``double exchange'' coming from the JT term. The 
effective hamiltonian is
\bq
\nonumber
H_{eff} & = & -t_{\uparrow}(M)\sum_{<ij>}(c_{i\uparrow}^{\dag} 
c_{j\uparrow}+h.c) - t_{\downarrow}(M)\sum_{<ij>}
(c_{i\downarrow}^{\dag}c_{j\downarrow}+h.c) \\ 
& + &
V\sum_{i}n_{i\uparrow}n_{i\downarrow} - g\sum_{i}{\bf \tau_{i}}.{\bf Q_{i}}
+k/2\sum_{i}Q_{i}^{2} \;,
\eq
with $t_{\alpha}(M)=t_{\alpha}[(1+M^{2})/2]^{1/2}$~\cite{[9]}.
In addition, interactions between lattice distortions at different sites is
will stabilize a particular distortion direction.  We model this effect by
introducing a term $h'=\kappa\sum_{<ij>\alpha}Q_{i}^{\alpha}Q_{j}^{\alpha}$
with $\kappa$ chosen to be ferromagnetic~\cite{[4]}. This is consistent 
with the observation in the FM metallic state, and consideration of 
staggered phases will require a two-sublattice extension.  We do not 
consider this in this work.

To simplify matters further, it is convenient to introduce a local pseudospin
quantization axis parallel to the local ${\bf Q_{i}}$ at each site.  This 
rotation from a fixed quantization axis to the local one corresponds to 
daigonalizing the JT part above by a local unitary transformation, $U_{i}$:
$U_{i}^{\dag}[({\bf \tau_{i}}.{\bf Q_{i}})/Q_{i}]U_{i}=\tau_{iz}$.  One has
simultaneously to transform the electron operators as, 
$a_{i\sigma}=U_{i}c_{i\sigma}$.  After doing the transformations, the 
effective hamiltonian takes the form,
\bq
\nonumber
H_{eff} & = & \sum_{<ij>,\sigma,\sigma'}t_{ij}({\bf S})U_{i}^{\dag} U_{j}
a_{i\sigma}^{\dag}a_{j\sigma'} + V\sum_{i}n_{i\uparrow}n_{i\downarrow} \\ 
& - & g\sum_{i\sigma}Q_{i}\sigma n_{i\sigma} + k/2\sum_{i}Q_{i}^{2} \;.
\eq
Notice that the JT term has been rendered simpler, but we have to pay the 
price by having to consider a complicated hopping term.  However, 
in the FM state around $x=0.3$, there is no orbital LRO, and at low-$T$, 
there is no evidence for localization effects.  The off-diagonal dependence 
of $t_{ij}$ is then not crucial in this regime, and so we replace 
$U_{i}^{\dag}t_{ij}U_{j}=t_{eff}$.  With this assumption, we have reduced 
our problem to that of a Hubbard model in orbital space, with an additional 
site-diagonal (random in the para-orbital state ($x=0.3$) ``disorder'' 
potential coming from the disordered (or short-range-ordered) JT distortions.
This term acts like a random magnetic field on the orbital pseudospins.  
Given this, the effective hamiltonian is solved within $d=\infty$ by 
{\it iterated perturbation theory}, extended to include effects arising from 
repeated scattering due to diagonal disorder~\cite{[18]}.
In what follows, we will use the extended IPT (IPT+CPA) to treat the combined
effects of orbital (Hubbard-like) correlations and static (JT) disorder in the
low-$T$ ferromagnetic metallic state around $x=0.3$.

We use the $d=\infty$ solution~\cite{[16]} of the Hubbard-like model 
without the
JT-term (since no JT distortion is observed in the FM state at $x=0.3$) to 
provide a simple understanding of various features observed in experiments.
Firstly, we notice that the anomalous stabilization of the CO state in 
external magnetic fields has a simple explanation:  $B_{ext}$ increases 
the effective hopping in the above eqn., leading to an increase in $T_{CO}$, 
via $T_{CO}\simeq (t^{2}\gamma^{2}(M)/V)$ for $V>>t$ (since the orbital AF 
state is the analog of the Neel-ordered AF state in the usual Hubbard model) 
as can be seen from the Hubbard model physics in $d=\infty$ on simple cubic  
lattice both in the weak and strong coupling regimes. If this low-$T$ 
transition is indeed driven by the development of orbital ordering,
via the above mechanism, $T_{CO}$ should show an isotope effect, which we 
quantify later. Clearly, this requires the orbital effects via $V$, and so 
arises from the correlation between orbital and magnetic degrees of freedom, 
as envisaged in~\cite{[12]}.  
In the ``overdoped'' samples ($x=0.33$), one is in 
the ``weak scattering'' regime, with a Fermi liquid response at low-$T$.  
The transfer of spectral weight with $T$, a characteristic feature of 
manganites, again has a natural interpretation in terms of Hubbard model 
physics with magnetization dependent hoppings, as in the above eqn.

Quite generally, the optical conductivity in $d=\infty$ is given by a simple
bubble diagram involving the full \mbox{$d=\infty$} local GF 
$G_{loc}(i\omega_{n})$:~\cite{[16]}

\bq
\nonumber
\sigma(\omega) & = & \frac{2e^{2}t_{\uparrow}^{2}(M)a^{2}}
{\Omega\hbar^{2}}\int_{-\infty}^{\infty}dED(E)\int_{-\infty}^{\infty}
\frac{d\omega'}{2\pi}<\rho(E,\omega')> \\
& \times &<\rho(E,\omega+\omega')>
\frac{n_{F}(\omega')-n_{F}(\omega'+\omega)}{\omega} \;,  
\eq
where $<\rho(E,\omega)>=-Im <[\omega-E-\Sigma(\omega)]^{-1}>/\pi$ is the 
disorder-averaged single-particle spectral fn. of the disordered Hubbard 
model.

\begin{figure}[htb]
\epsfxsize=3.4in
\epsffile{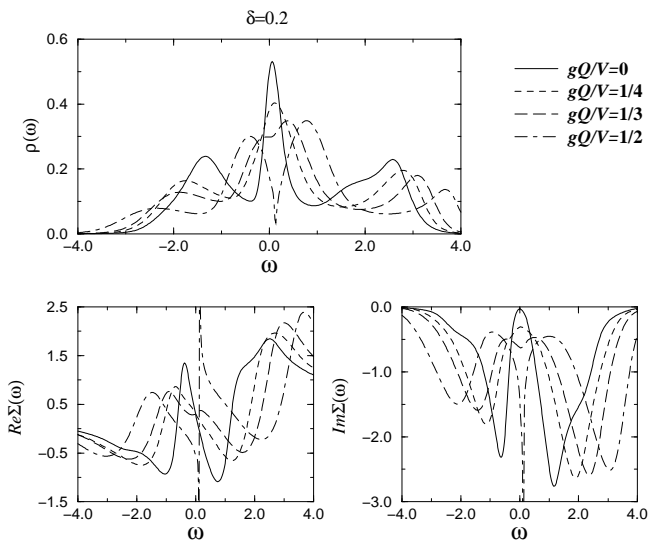}
\caption{ Local spectral density and the real and imaginary parts of the 
s.p self energy for the disordered, orbital Hubbard model in the paper in 
$d=\infty$  at $T=0.01D$.  Notice the sharp FL feature without JT disorder, 
which is broadened out for small $Q_{i}$ while preserving the FL picture 
(dashed line). This corresponds to the FM metallic phase of the manganites.  
With increasing $Q_{i}$, a pseudogap opens up in $\rho(\omega)$, and the 
scattering rate describes an incoherent non-FL metal.} 
\label{fig1}
\end{figure}

In fig.~(1), we show the density of states and the self-energy for the model
in $d=\infty$.  We choose $x=n=0.8$, $U/D=3.0$ and $\lambda/D=0.2$.
These parameters are realistic if $D$ is chosen to be around $1 eV$.  
We also choose a simple binary distribution for the JT disorder; this is 
motivated by the fact that the $Mn^{4+}$ octahedra are JT-undistorted- only 
the $Mn^{3+}$ are JT-distorted.  This effective disorder is modelled by 
$P(Q_{i})=(1-x)\delta(Q_{i})+x\delta(Q_{i}-Q)$.  We restrict ourselves to 
the low-$T$ state, and so choose $T=0.01D$.  Notice that 
$Im\Sigma(\omega) \simeq -C-D\omega^{2}$, in good accord with the results 
extracted from the optical data of Simpson {\it et al}~\cite{[8]}.
The carrier scattering rate thus goes quadratically with $\omega, T$, as seems 
to be observed in the FM metallic state.  Simultaneously, $Re \Sigma(\omega) 
\simeq -a\omega$ at low energy, and the quasiparticle picture is valid inspite 
of JT disorder.  For comparison, we also show the results without disorder; in
this case, our results are in good agreement with those from earlier 
studies~\cite{[16]}; this serves as a check on our numerics.  
The longitudinal optical conductivity, computed using the above eqn, 
also reveal features in good agreement with those observed in experiments.   
As $T$ is lowered, the FM spin polarization increases, leading 
to increase in $t_{1,2}$ and, within Hubbard model physics in $d=\infty$, to a 
transfer of spectral weight on a scale of $V$ from the high-energy incoherent 
part to a low energy quasicoherent peak.  Furthermore, the optical 
conductivity experiments reveal that the low energy spectral weight scales 
like $\gamma^{2}(M)=(1+M^{2})/2$ at low-$T$ (but not above $T_{c}^{FM}$), 
exactly as is expected from the effective hamiltonian above.  In the low-$T$ 
FM metallic phase, we see clearly the Drude response at low energy, followed 
by a broad hump at higher energy; we attribute this to the orbital 
correlations in the $e_{g}$ sector    ($U$) in the model above.  Notice 
that in presence of strong orbital correlations, the spectral weight 
transfer from high to low energies is controlled by the ratio $V/t_{ab}(M)$ 
(which is $T$-dependent in our model, via the $T$-dependence of $M$); as 
$T$ decreases, $M(T)$ and hence $t_{ab}(M)$ increases,
reducing the ratio $V/t_{ab}(M)$ and leading to a transfer of spectral weight 
from high- to low energies, as observed. While the  spectral weight transfer 
with $T$ can be rationalized by simple DE models~\cite{[2]}, the small weight 
of the quasicoherent Drude peak~\cite{[8]}, as well as the intense 
mid-infrared absorption is clearly unexplainable in such approaches, but has 
a natural interpretation within our picture, where it arises from the 
transition from the ``lower Hubbard subband'' (in the doubly degenerate 
$e_{g}$ sector) to the quasicoherent Kondo like peak, as in the usual 
Hubbard model in $d=\infty$~\cite{[16]}. 

Fig.(2) indeed shows these features; the JT "disorder" broadens the Drude 
peak somewhat, yielding the constant $C$ in the scattering rate, while 
leaving the $\omega^{2}$ part unchanged.  The mid-IR peak, corresponding 
to transitions between the lower-Hubbard band and the quasiparticle 
resonance in the $d=\infty$  Hubbard model, is shifted somewhat, to 
$\omega'=V/2+gQ$, and the high-energy feature (completely broadened out) 
corresponds to transitions between the Hubbard bands.  Our computed result 
for the optical conductivity with $gQ/V=1/4$ is in good agreement with 
the thin film results of Simpson {\it et al.}~\cite{[8]}.  With increasing 
$gQ/V$ (fig.2), we see that the optical response at low energies is totally 
incoherent, as was indeed observed in earlier studies~\cite{[6]}.  In our 
picture, this is simply a consequence of the increased strong scattering 
from JT- as well as doping induced disorder, which produces an incoherent 
spectral density and a non-Drude optical response. 
The s.p scattering rate is related to
$Im\Sigma(\omega) \simeq 
-(V^{2}/D_{\uparrow}^{2}D_{\downarrow})\omega^{2}n_{\downarrow}(1-n_{\downarrow}
)$, 
within second-order perturbation theory (this actually gives the correct FL
estimate at low energy in $d=\infty$). This implies a real part that goes
like $Re 
\Sigma(\omega)=-(V^{2}/D_{\uparrow}^{2})n_{\downarrow}(1-n_{\downarrow})\omega$, 
and
\begin{figure}[hbt]
\epsfxsize=3.4in
\epsffile{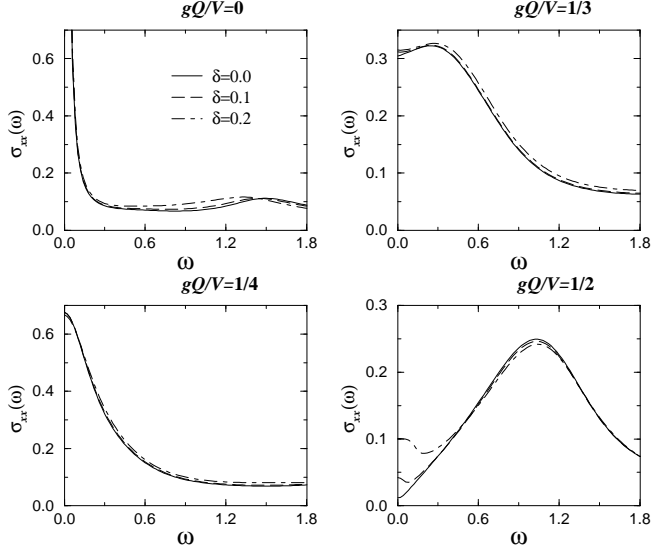}
\caption{Optical conductivity, $\sigma_{xx}(\omega)$ of the disordered, 
orbital Hubbard model in $d=\infty$ as a fn. of doping for various 
values of the JT disorder, $gQ/V=0,~1/4,~1/3,~1/2$.  The case $gQ/V=1/4$ 
corresponds most closely to the conditions of 
ref.~\cite{[8]}. Notice the 
somewhat broadened, but still dominant, Drude response at low energy, 
followed by a mid-IR hump that carries a large fraction of the total 
optical weight.  The other curves may correspond to situations with larger 
ratio $gQ/V$, as in situations where small or no Drude response is seen.} 
\label{fig2}
\end{figure}
\hspace{-.5cm} 
hence the quasiparticle residue at $\mu$ is 
$Z=(1+(V/D)^{2}n_{\downarrow}(1-n_{\downarrow}))^{-1}$.  So $m^{*}/m \simeq 3 $ 
for $V=3D$ and 
$x=0.3$, giving a specific heat enhancement in the observed range. Notice 
that this is also close to the optical mass extracted by Simpson {\it et al.} 
from their optical measurements, confirming that charge (orbital) ordering 
tendencies are not important deep in the FM metallic state. The above 
calculation also shows that the resistivity, $\rho(T) \simeq AT^{2}$ with 
$A=(V^{2}/D_{\uparrow}^{2}(M)D_{\downarrow}(M))$ decreasing with increasing 
$B_{ext}$, as observed.  The other terms correspond to elastic scattering 
off the core-spin and the inelastic scattering off two-magnon fluctuations, 
and are given by $\rho_{m}(T) \simeq \rho_{0}+BT^{9/2}$ as shown by Kubo.
Since the scattering off spin as well as orbital fluctuations correspond to 
single-site processes in $d=\infty$, it follows that the total scattering
rate is the sum of those coming from magnetic and orbital scattering.  This 
gives a consistent explanation of $\rho(T)$ below $T_{c}^{FM}$.  More 
importantly, it is also consistent with results from Raman scattering 
experiments, as seen from the following argument.  In $d=\infty$, the 
vertex corrections do not enter the conductivity~\cite{[16]}; in this 
approximation the Raman intensity is simply related to the conductivity as 
long as the incident laser frequency is not close to those corresponding 
to resonant scattering.  The Raman intensity is given by~\cite{[19]},
\be
I(\omega)=\frac{\omega}{1-e^{-\beta\omega}}\sigma_{xx}'(\omega) \;.
\ee
Using the Drude form for $\sigma_{xx}(\omega)$ in the para-orbital FL regime,
$\sigma_{xx}(\omega)=\frac{\omega_{p}^{2}/4\pi}{1/\tau-i\omega}$, with 
$1/\tau \simeq x Im\Sigma(\omega,T) + \Gamma_{0}$, where $\Gamma_{0}$ is the
impurity contribution to the scattering rate (from residual disorder) gives, 
for $\omega << k_{B}T$,
\be
I(\omega)=\frac{\omega_{p}^{2}}{4\pi}k_{B}T\frac{1}{\Gamma_{0}+xA(M)
\omega^{2}-i\omega} \;,
\ee
with $\alpha=xA(M)=xV^{2}/D_{\uparrow}^{2}(M)D_{\downarrow}(M)$. This 
explains the coherent $Im \Sigma(\omega) = \Gamma_{0}+\alpha\omega^{2}$ 
term needed to understand Raman lineshapes below $T_{c}^{FM}$. Since the 
selfenergy is purely local in $d=\infty$, the Fermi surface is unchanged 
(from its bandstructure value) by interactions in the para-orbital metallic 
state.  The discontinuity in $n({\bf k})$, the momentum occupation fn, 
is reduced by interactions, and is quite small for $V=3D$ 
(off $n=1$)~\cite{[16]}, completely consistent with the small
discontinuity observed in the FM metallic state at $x=0.3$.  The Fermi 
surface is large, as expected from bandstructure calculations, and in 
agreement with the small value of the Hall constant.  Notice that 
calculation of the normal contribution to the Hall coefficient in the 
$d=\infty$ Hubbard model~\cite{[20]} would yield a small value of 
$R_{H}(x)$ for $x \simeq 0.3$, consistent with the above. Since the 
bandwidth $D$ increases in an external magnetic field via $\gamma(M)$, 
the coefficient $A$ decreases, and we predict enhancement of the linear 
specific heat coefficient with $B_{ext}$, as well as a decreased value 
of $R_{H}$, which seem to be observed in the FM metallic state.

The dynamical effects of the JT term can now be considered. 
At low-$T$, the effects of electron-lattice interaction can be absorbed into
the effective $V$ in our orbital Hubbard model.  Given an optical phonon 
frequency $\omega_{0}$, at temperatures $T<<\hbar\omega_{0}/k_{B}$, one 
can integrate out the phonons, generating a term 
$H' \simeq \frac {\lambda^{2}}{\hbar\omega_{0}}\sum_{i}[c_{i\alpha}^{\dag}({\bf 
\tau}/2)c_{i\beta}]^{2}$.  At low-$T$, with $J_{H} \rightarrow \infty$, 
this looks like an additional Hubbard interaction in orbital space, and so 
the results obtained above are qualitatively unaffected except for the change 
$V \rightarrow V+\frac {\lambda^{2}}{\hbar\omega_{0}}$.
Given a larger $V_{eff}/D$ in our orbital Hubbard model, the consequence 
within DMFT will be to decrease the coherent spectral weight, moving it to the 
incoherent, high-energy region.  Thus, the $A$ in the quadratic term in the 
resistivity will be $A=V_{eff}^{2}/D_{\uparrow}^{2}D_{\downarrow}$, and the
Drude-like component in the optical conductivity is further suppressed.  At
higher $T$, a description involving static JT distortions should 
suffice~\cite{[4]}.Finally, since 
$T_{c}^{FM} \simeq D^{*}=ZD(\omega_{0})$, where $D(\omega_{0})=D 
e^{-\lambda^{2}/\omega_{0}^{2}}$ is the polaron reduction of the 
bandwidth~\cite{[4]} in 
the DE model with $J_{H} \rightarrow \infty$, the scaling of the FM transition
temperature with ionic mass is given by
\be
k_{B}T_{c}^{FM}=\frac{D}
{1+((V+\lambda^{2}/\hbar\omega_{0})/D)^{2}n_{\downarrow} (1-n_{\downarrow})}\;,
\ee
with $\omega_{0}^{2}= K/M_{O}$.  This represents a different way of
 understanding the isotope effect in $T_{c}^{FM}$ as arising from two 
effects; a polaron induced narrowing of the bandwidth, and a renormalization 
of the effective $V$ by carrier-phonon interactions.  Notice that such an 
explicit dependence on $\omega_{0}$ is absent in the adiabatic treatment of 
the phonons (in agreement with R\"oder {\it et al.}~\cite{[4]}.

  Finally, the orbital-charge ordering in the $x=1/8$ sample should show an 
isotope effect via $k_{B}T_{CO} \simeq (t^{2}\gamma^{2}(M)/V_{eff})$; i.e, 
a reduced isotopic mass $M$ will increase $\omega_{0}$ and decrease $V_{eff}$
thus stabilizing the $CO$ ordered state.  We are not aware of experiments 
addressing this issue, but we point out that it would be a nice confirmation 
of the importance of (strong) orbital correlations + JT effects in the CMR
manganites.

  Consideration of the high-$T$ "insulating" phase requires more work;  to see
this, notice that with strong $gQ > D$, the combined effects of disorder
and interactions in $H_{eff}$ will be to localize the carriers.  The decrease 
in the effective bandwidth (via decreasing $\gamma(M)$ above $T_{c}^{FM}$
favors carrier localization, and the enhancement of $D$ below $T_{c}^{FM}$
delocalizes the carriers (by moving the effective $\lambda/D$ ratio through a 
critical value required for carrier delocalization.  An external magnetic 
field has precisely this effect; the increase of $D(M)$ suppresses the 
high-$T$ insulating state, leading to the CMR effect.  The consideration of 
localization effects crucial for such a scenario is outside the scope of the 
$d=\infty$ approach~\cite{[16]}, and, indeed, we shall return to this 
question in a separate work.  Consideration of dynamical charge and spin 
correlations and their effects on magnetotransport near the various 
charge-orbital ordered phases in the global phase diagram requires a 
consideration of realistic hopping matrix elements, and more importantly, 
the inclusion of short-ranged (non-local) dynamical correlations in a proper 
extension of the DMFT (e.g, via the dynamical cluster approximation 
(DCA)~\cite{[21]}).  We plan to report such studies in a future publication.

To summarize, the low-$T$ features in the well doped FM metallic state of 
the manganites can be understood in a unified way based on the exact 
$d=\infty$ solution of the strong coupling ($U, J_{H} \rightarrow \infty$) 
double exchange model with inclusion of the orbital degeneracy and 
Jahn-Teller effects.  We have shown that orbital fluctuations play a very 
important role in the understanding of key features of the dc and ac 
transport even in the FM metallic state.  Specifically, orbital 
correlations generate a low-energy coherence scale that is sensitive to 
external magnetic fields. Additionally, these correlations narrow down the 
bandwidth, making it easier for a combination of hopping and structural 
disorder to induce carrier localization.  Recent work by Held {\it et 
al.}~\cite{[22]} considers the effect of the orbital $U$ in the paramagnetic 
phase of the
manganites.  However, the effects of the disordered Jahn-Teller coupling has
not been considered there.  In contrast, we have considered these effects 
along with strong interorbital correlations on the same footing, and have 
made a somewhat detailed comparison with experimental results in the low-$T$
FM state.  Our work is thus complementary to theirs. 

Recently, Nagaosa {\it et al.}~\cite{[22]} have shown 
that a slightly extended version of the model considered here (including super
exchange terms) and realistic hopping matrix elements gives a good agreement
with the experimentally observed phase diagram over the whole range of $x$.
We expect our simplified model (neglecting superexchange) to be an adequate
starting point to describe the FM metallic state at $x=0.3$.  It can be shown 
that inclusion of realistic hopping matrix elements will not qualitatively
affect the conclusions obtained above in $d=\infty$; they will, however, be
crucial when one attempts to understand the phase diagram, which shows a rich
sequence of transitions as a function of $x$.  We have not attempted to do 
this here; a more detailed numerical investigation to consider these 
additional details is under progress and will be reported separately.

\acknowledgments
One of us (MSL) acknowledges financial support of SFB 341.
LC was supported by the Funda\c c\~ao de Amparo \`a Pesquisa do Estado 
de S\~ao Paulo (FAPESP) and the Max-Planck Institut f\"ur Physik komplexer 
Systeme.

\end{document}